\documentclass{PoS}
\usepackage{graphicx}
\usepackage{amsfonts}
\usepackage{amssymb}
\usepackage{amsbsy}
\usepackage{amsmath}
\usepackage{latexsym}
\usepackage{bm}

\newcommand*{\di}{\partial}

\renewcommand*{\c}{\text{c}}

\newcommand*{\GR}{\text{\tiny{GR}}}
\newcommand*{\KK}{\mathcal{E}}

\renewcommand*{\b}{\text{b}}

\newcommand*{\approaches}[2]{\xrightarrow[#2]{\,\,\,{#1}\,\,\,}}

\newcommand*{\eff}{\text{\tiny{eff}}}
\newcommand*{\fiveD}{\text{\tiny{5D}}}
\newcommand*{\prim}{\text{\tiny{inf}}}

\title{Scalar perturbations in Randall-Sundrum braneworld cosmology}

\ShortTitle{Scalar perturbations in Randall-Sundrum braneworld
cosmology}

\author{\speaker{Antonio CARDOSO}%
        \thanks{Based on the work done in collaboration with Takashi Hiramatsu, Kazuya Koyama and Sanjeev S. Seahra, reported in \cite{Cardoso:2007zh}.}\\
       Institute of Cosmology \& Gravitation, University of Portsmouth, Portsmouth~PO1~2EG, UK\\
       E-mail: \email{Antonio.Cardoso@port.ac.uk}}

\abstract{We study the evolution of scalar perturbations in the
radiation-dominated era of Randall-Sundrum braneworld cosmology by
numerically solving the coupled bulk and brane master wave
equations. We find that density perturbations with wavelengths less
than a critical value (set by the bulk curvature length) are
amplified during horizon re-entry. Conversely, we explicitly confirm
from simulations that the spectrum is identical to GR on large
scales. Although this magnification is not relevant for the cosmic
microwave background or measurements of large scale structure, it
may have some bearing on the formation of primordial black holes in
Randall-Sundrum models.}

\FullConference{Cargèse Summer School: Cosmology and Particle Physics Beyond the Standard Models\\
         July 30 - August 11, 2007\\
         Institut d'Etudes Scientifiques de Cargese}

\begin{document}

\section{Introduction}

The Randall-Sundrum (RS) braneworld model \cite{Randall:1999vf}
postulates that our observable universe is a thin 4-dimensional
hypersurface residing in 5-dimensional anti-de Sitter (AdS) space.
Ordinary matter degrees of freedom are assumed to be confined to the
brane, while gravitational degrees of freedom are allowed to
propagate in the full 5-dimensional bulk. The warping of AdS space
allows us to recover ordinary general relativity (GR) at distances
greater than the curvature radius of the bulk $\ell$. Current
laboratory tests of Newton's law constrain $\ell$ to be less than
around $0.1\,\text{mm}$ \cite{Kapner:2006si}.

It is well known that the Friedmann equation governing the expansion
of the brane universe differs from general relativity by a
correction of order $\rho/\sigma$, where $\rho$ is the energy
density of the brane matter and $\sigma \gtrsim (\text{TeV})^4$ is
the brane tension. The magnitude of this correction defines the
``high-energy'' regime of braneworld cosmology as the era when $\rho
\gtrsim \sigma$ or equivalently $H\ell \gtrsim 1$, where $H$ is the
Hubble parameter.

The equations of motion governing fluctuations of the model are
found to differ from GR in two principal ways at early times: First,
they acquire $\mathcal{O}(\rho/\sigma)$ high-energy corrections
similar to those found in the Friedmann equation. By themselves,
such corrections are not difficult to deal with: they just modify
the second-order ordinary differential equations (ODEs) governing
perturbations in GR. But the second type of modification is more
problematic: perturbations on the brane are also coupled to
fluctuations of the 5-dimensional bulk geometry, which are
collectively known as the ``Kaluza Klein'' (KK) degrees of freedom
of the model. The KK modes are governed by master partial
differential equations (PDEs) defined throughout the AdS bulk
\cite{Mukohyama:2000ui,Kodama:2000fa}. The only known way of solving
this system of equations on all scales simultaneously is by direct
numerical solution.

The purpose of the paper is to numerically solve for the behaviour
of scalar perturbations in the radiation-dominated regime of
braneworld cosmology. We use two different numerical codes recently
developed in Refs.~\cite{Hiramatsu:2006cv,Cardoso:2006nh}, which
gives us the ability to confirm the consistency of our numeric
results via two independent algorithms. We are ultimately interested
in finding the matter transfer function in the radiation era, and
also determining the relative influence of KK and high-energy
effects on the density perturbations. Heuristically, we may expect
the KK modes to amplify high-energy/small-scale density
perturbations. The reason is that we know that the gravitational
force of attraction in the RS model is stronger than in GR on scales
less than $\ell$ \cite{Randall:1999vf,Garriga:1999yh}. This implies
that modes with a physical wavelength smaller than $\ell$ during
horizon crossing will be amplified due to the KK enhancement of the
gravitational force. However, this physical reasoning needs to be
tested with numeric simulations.

\section{Scalar perturbations}

It has been shown in Refs.~\cite{Mukohyama:2000ui,Kodama:2000fa}
that scalar-type perturbations of the bulk geometry are governed by
a single gauge invariant master variable $\Omega$. This bulk master
variable satisfies the following wave equation
\begin{equation}\label{eq:bulk wave equation}
    0 = -\frac{\di^2\Omega}{\di \tau^2} + \frac{\di^2\Omega}{\di z^2}
    + \frac{3}{z} \frac{\di\Omega}{\di z} + \left( \frac{1}{z^2} -
    k^2 \right) \Omega,
\end{equation}
and a boundary condition on the brane
\begin{equation}\label{eq:boundary condition}
    \left[ \di_n \Omega + \frac{1}{\ell} \left(1 +
    \frac{\rho}{\sigma} \right) \Omega + \frac{6\rho a^3}{\sigma
    k^2} \Delta \right]_\b = 0.
\end{equation}
The density contrast on the brane $\Delta$ satisfies the wave
equation
\begin{subequations}\label{eq:brane wave equation}
\begin{gather}
    \frac{d^2 \Delta}{d\eta^2} + (1+3c_s^2-6w) Ha \frac{d\Delta}{d\eta} +
    \left[c_s^2 k^2 + \frac{3\rho a^2}{\sigma\ell^2} A +
    \frac{3\rho^2 a^2}{\sigma^2\ell^2} B \right] \Delta = -\frac{k^2\Gamma}{\rho} + \frac{k^4(1+w)\Omega_\b}{3\ell a^3}, \\
    A = 6c_s^2 -1 -8w+3w^2, \qquad
    B = 3c_s^2-9w-4.
\end{gather}
\end{subequations}
Here, we have defined the sound speed $c_s^2 = \delta p /\delta
\rho$ and the equation of state $w=p/\rho$ (where $\rho$ and $p$ are
the energy density and pressure of the brane fluid, respectively),
$\Gamma$ is the entropy perturbation of the brane matter and $\eta$
is the conformal time along the brane. In this paper we assume that
the matter anisotropic stress vanishes. The above ODE, the bulk wave
equation (\ref{eq:bulk wave equation}) and the boundary condition
(\ref{eq:boundary condition}) comprise a closed set of equations for
$\Delta$ and $\Omega$. Note that in the low energy universe, we can
neglect $\mathcal{O}(\rho^2/\sigma^2)$ terms. If we also set
$\Omega_\b = 0$ we obtain the standard 4-dimensional dynamical
equation for $\Delta$; hence, we recover GR at low energies.

\section{Numeric analysis and discussion}

For the rest of the paper, we will restrict ourselves to the case of
a radiation-dominated brane with $w = 1/3$. We define the ``*''
epoch as the moment in time when a mode with wavenumber $k$ enters
the Hubble horizon, $k = H_* a_*$. Another important era is the
critical epoch, the transition between high and low energy regimes,
when $H_c\ell = 1$ and the radiation density has its critical value
$\rho_c/\sigma = \sqrt{2}-1$. Generally speaking, we call modes with
$k > k_\c$ ``supercritical'' and modes with $k<k_\c$
``subcritical''.  The scale defined by the critical mode in today's
universe corresponds, for $\ell = 0.1\,\text{mm}$, to a scale of
$\sim 10$ astronomical units (AU), which is incredibly tiny by
cosmological standards.

\begin{figure}
\begin{center}
    \includegraphics[width=14.5cm]{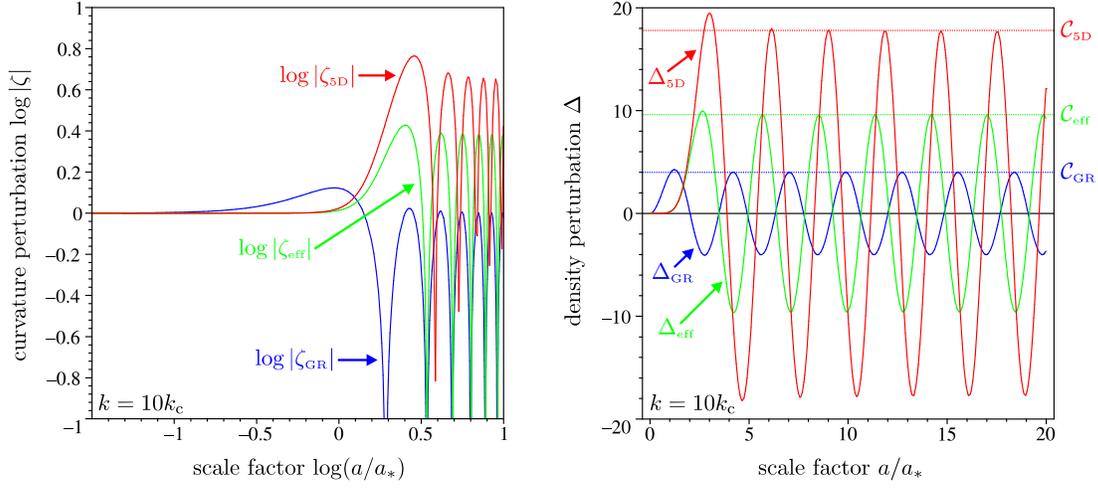} \caption{A comparison of the
    behaviour of the curvature perturbation $\zeta$ (\emph{left}) and the density
    perturbation $\Delta$ (\emph{right}) in the full 5-dimensional theory including
    KK contributions (5D), the effective 4-dimension theory including $\mathcal{O}(\rho/\sigma)$
    corrections (eff), and ordinary general relativity (GR).  The waveforms for each theory are normalized
    such that $\zeta = 1$ on superhorizon scales.\label{fig:effective}}
\end{center}
\end{figure}

In Fig.~\ref{fig:effective}, we plot the predictions of GR, the
4-dimensional effective theory (where all $\mathcal{O}(\rho/\sigma)$
corrections to GR are retained, but the bulk effects are removed by
artificially setting $\Omega = 0$), and the full 5-dimensional
simulations for the behaviour of the curvature perturbation on
uniform density slices $\zeta$ and the density contrast $\Delta$ for
a supercritical mode. Since in any given model we expect the
primordial value of the curvature perturbation to be fixed by
inflation, it makes physical sense to normalize the waveforms from
each theory such that $\zeta_\fiveD \approx \zeta_\eff \approx
\zeta_\GR \approx 1$ for $a \ll a_*$. When this is enforced we see
that the effective theory predicts a larger final amplitude for the
density perturbation than GR. Furthermore, the final amplitude in
the 5-dimensional simulation is larger than both of the other
theories.  From this we can infer that, as we expected, both
$\mathcal{O}(\rho/\sigma)$ and KK effects induce enhancement in the
amplitude of perturbations.

\begin{figure}
\begin{center}
    \includegraphics[width=14.5cm]{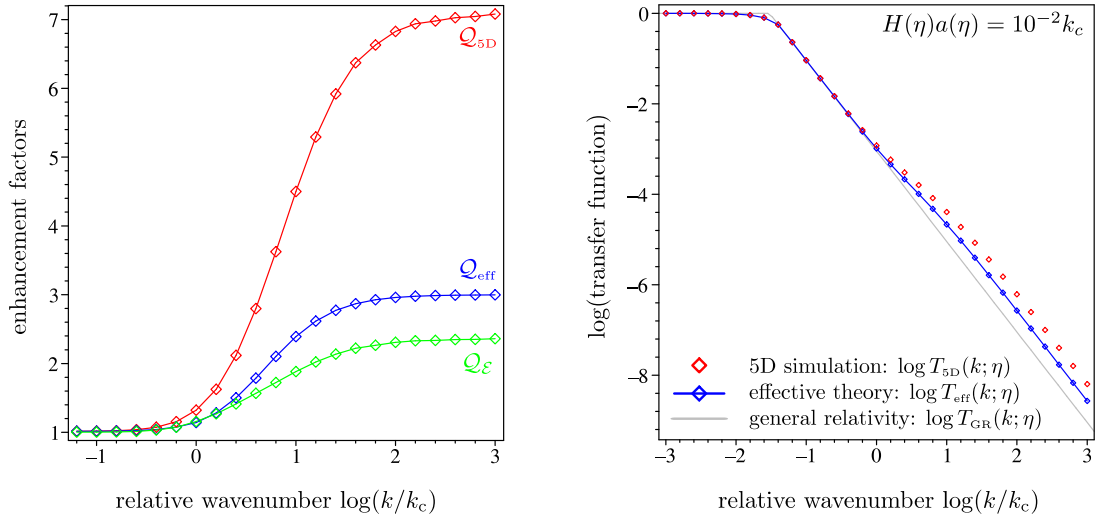} \caption{Density perturbation
    enhancement factors (\emph{left}) and transfer functions
    (\emph{right}) from simulations, effective theory, and general
    relativity. The transfer functions in
    the right panel are evaluated at a given subcritical epoch in the
    radiation dominated era.\label{fig:spectra}}
\end{center}
\end{figure}

As in Fig.~\ref{fig:effective}, let the final amplitudes of the
density perturbation with wavenumber $k$ be $\mathcal{C}_\fiveD(k)$,
$\mathcal{C}_\eff(k)$ and $\mathcal{C}_\GR(k)$ for the
5-dimensional, effective and GR theories, respectively. Then, we
define enhancement factors as
\begin{equation}
    \mathcal{Q}_\eff(k) =
    \frac{\mathcal{C}_\eff(k)}{\mathcal{C}_\GR(k)}, \quad \mathcal{Q}_\KK(k) =
    \frac{\mathcal{C}_\fiveD(k)}{\mathcal{C}_\eff(k)}, \quad \mathcal{Q}_\fiveD(k) =
    \frac{\mathcal{C}_\fiveD(k)}{\mathcal{C}_\GR(k)}.
\end{equation}
It follows that $\mathcal{Q}_\eff(k)$ represents the
$\mathcal{O}(\rho/\sigma)$ enhancement to the density perturbation,
$\mathcal{Q}_\KK(k)$ gives the magnification due to KK modes, while
$\mathcal{Q}_\fiveD(k)$ gives the total 5-dimensional amplification
over the GR case. These enhancement factors are shown in the left
panel of Fig.~\ref{fig:spectra}.  We can see that they all increase
as the scale is decreased, and that they all approach unity for $k
\rightarrow 0$, which means we recover general relativity on large
scales.  For all wavenumbers we see $\mathcal{Q}_\eff
> \mathcal{Q}_\KK > 1$, which implies that the amplitude magnification due
to the $\mathcal{O}(\rho/\sigma)$ corrections is always larger than
that due to the KK modes.  Interestingly, the $\mathcal{Q}$-factors
appear to approach asymptotically constant values for large $k$.

Now we consider a transfer function $T(k)$ that will tell us how the
initial spectrum of curvature perturbations
$\mathcal{P}_\zeta^\prim$ maps onto the spectrum of density
perturbations $\mathcal{P}_\Delta$ at some low energy epoch within
the radiation era characterized by the conformal time $\eta >
\eta_\c$. It is customary to normalize transfer functions such that
$T(k;\eta) \approaches{k}{0} 1$, which leads us to the following
definition
\begin{equation}
    T(k;\eta) = \frac{9}{4} \left[ \frac{k}{H(\eta)a(\eta)}
    \right]^{-2} \frac{\Delta_k(\eta)}{\zeta^\prim_k}.
\end{equation}
Here, $\zeta^\prim_k$ is the primordial value of the curvature
perturbation and $\Delta_k(\eta)$ is the maximum amplitude of the
density perturbation in the epoch of interest. We know that we
recover the GR result in the extreme small scale limit, which gives
the transfer function the correct normalization. In the righthand
panel of Fig.~\ref{fig:spectra}, we show the transfer functions
derived from GR, the effective theory and the 5-dimensional
simulations.  As expected, the $T(k;\eta)$ for each formulation
match one another on subcritical scales $k < k_\c$. However, on
supercritical scales we have $T_\fiveD > T_\eff > T_\GR$. Our
results are robust against modifications of the initial data for
simulations.

The amplitude enhancement of perturbations is important on comoving
scales $\lesssim 10 \,\text{AU}$, which are far too small to be
relevant to present-day/cosmic microwave background measurements of
the matter power spectrum. However, it may have an important bearing
on the formation of compact objects such as primordial black holes
and boson stars at very high energies, i.e. the greater
gravitational force of attraction in the early universe will create
more of these objects than in GR. We know that the abundance of
primordial black holes can be constrained by big bang nucleosythesis
and observations of high-energy cosmic rays, so it would be
interesting to see if the kind of enhancement of density
perturbations predicted in this paper can be used to derive new
limits on Randall-Sundrum cosmology.

\begin{acknowledgments}
I am supported by FCT (Portugal) PhD fellowship SFRH/BD/19853/2004.
\end{acknowledgments}


\begin{thebibliography}{99}

%\cite{Cardoso:2007zh}
\bibitem{Cardoso:2007zh}
  A.~Cardoso, T.~Hiramatsu, K.~Koyama and S.~S.~Seahra,
  %``Scalar perturbations in braneworld cosmology,''
  arXiv:0705.1685 [astro-ph].
  %%CITATION = ARXIV:0705.1685;%%

%\cite{Randall:1999vf}
\bibitem{Randall:1999vf}
  L.~Randall and R.~Sundrum,
  %``An alternative to compactification,''
  Phys.\ Rev.\ Lett.\  {\bf 83} (1999) 4690
  [arXiv:hep-th/9906064].
  %%CITATION = PRLTA,83,4690;%%

%\cite{Kapner:2006si}
\bibitem{Kapner:2006si}
  D.~J.~Kapner, T.~S.~Cook, E.~G.~Adelberger, J.~H.~Gundlach, B.~R.~Heckel, C.~D.~Hoyle and H.~E.~Swanson,
  %``Tests of the gravitational inverse-square law below the dark-energy  length
  %scale,''
  Phys.\ Rev.\ Lett.\  {\bf 98} (2007) 021101
  [arXiv:hep-ph/0611184].
  %%CITATION = PRLTA,98,021101;%%

%\cite{Mukohyama:2000ui}
\bibitem{Mukohyama:2000ui}
  S.~Mukohyama,
  %``Gauge-invariant gravitational perturbations of maximally symmetric
  %spacetimes,''
  Phys.\ Rev.\  D {\bf 62} (2000) 084015
  [arXiv:hep-th/0004067].
  %%CITATION = PHRVA,D62,084015;%%

%\cite{Kodama:2000fa}
\bibitem{Kodama:2000fa}
  H.~Kodama, A.~Ishibashi and O.~Seto,
  %``Brane world cosmology: Gauge-invariant formalism for perturbation,''
  Phys.\ Rev.\  D {\bf 62} (2000) 064022
  [arXiv:hep-th/0004160].
  %%CITATION = PHRVA,D62,064022;%%

%\cite{Hiramatsu:2006cv}
\bibitem{Hiramatsu:2006cv}
  T.~Hiramatsu and K.~Koyama,
  %``Evolution of curvature perturbations in a brane-world inflation at
  %high-energies,''
  JCAP {\bf 0612} (2006) 009
  [arXiv:hep-th/0607068].
  %%CITATION = JCAPA,0612,009;%%

%\cite{Cardoso:2006nh}
\bibitem{Cardoso:2006nh}
  A.~Cardoso, K.~Koyama, A.~Mennim, S.~S.~Seahra and D.~Wands,
  %``Coupled bulk and brane fields about a de Sitter brane,''
  Phys.\ Rev.\  D {\bf 75} (2007) 084002
  [arXiv:hep-th/0612202].
  %%CITATION = PHRVA,D75,084002;%%

%\cite{Garriga:1999yh}
\bibitem{Garriga:1999yh}
  J.~Garriga and T.~Tanaka,
  %``Gravity in the brane-world,''
  Phys.\ Rev.\ Lett.\  {\bf 84} (2000) 2778
  [arXiv:hep-th/9911055].
  %%CITATION = PRLTA,84,2778;%%

\end{thebibliography}
\end{document}